\documentclass[aps,prd,onecolumn,notitlepage,nofootinbib,preprintnumbers]{revtex4}
\pdfoutput=1 
\usepackage{amsmath,latexsym,amssymb,graphicx,slashed,hyperref,color,enumerate,url,etoolbox}
\hypersetup{colorlinks,citecolor= nicegreen,linkcolor= nicered}
\definecolor{nicered}{rgb}{0.7,0.1,0.1}
\definecolor{nicegreen}{rgb}{0.1,0.5,0.1}
\begin{document}

\title{Self-interacting Dark Matter Without Direct Detection Constraints}
\author{Yue Zhang}
\affiliation{Department of Physics and Astronomy, Northwestern University, Evanston, IL 60208, USA}

\begin{abstract}
\noindent
We explore the self-interacting dark matter scenario in a simple dark sector model where the dark matter interacts through a dark photon. Splitting a Dirac fermion dark matter into two levels using a small Majorana mass can evade strong direct detection constraints on the kinetic mixing between the dark and normal photons, thus allowing the dark sector to be more visible at high intensity and/or high energy experiments. It is pointed out that such a mass splitting has a strong impact on the dark matter self-interaction strength. We derive the new parameter space of a pseudo-Dirac self-interacting dark matter. Interestingly, with increasing mass splitting, a weak scale dark matter mass window survives that could be probed by the LHC and future colliders.
\end{abstract}

\preprint{NUHEP-TH/16-06}

\maketitle

\section{Introduction}

In spite of the well established evidence for its existence in the universe, the nature of dark matter remains elusive. At present, most of the survived hints for particular properties of dark matter seem to come from the cosmological and astrophysical side. In the case the dark matter or part of it has a particle physics origin, it would be of great importance to probe the theory behind in the laboratories, just like probing the other known particles.

In this paper, we focus on a very simple setup where the dark matter is a fermion charged under a dark $U(1)$ gauge symmetry and its interactions are mediated by a massive vector gauge boson (dark photon) --- a massive dark QED model. Both the dark matter and the dark photon are singlets under the standard model gauge symmetries. The dark sector communicates with the standard model sector through the kinetic mixing term between the dark photon and the normal photon. Th model is very simple and allows one to do concrete calculations of many observable quantities about the dark matter. Over the recent years, it has been extensively studied in the literature for understanding various phenomena and experimental hints of dark matter, from the underground to the cosmos~\cite{TuckerSmith:2001hy, Finkbeiner:2007kk, ArkaniHamed:2008qn, Fox:2008kb, An:2016gad}. It also serves as an alternative to the supersymmetric WIMP paradigm and can accommodate much more freedom in the relative importance of direct and indirect detections~\cite{Pospelov:2007mp}. More recently, it has been shown that the ``long'' range force nature due to a light dark photon exchange enables the dark matter to have large enough self-scattering cross section~\cite{Ackerman:mha, Feng:2009mn, Buckley:2009in, Loeb:2010gj, Tulin:2012wi, Tulin:2013teo} to be a self-interacting dark matter (SIDM) candidate~\cite{Spergel:1999mh}. This allows it to provide an explanation to a few puzzles in the small scale structure formation such as the dwarf galaxy core/cusp and too-big-to-fail problems~\cite{BoylanKolchin:2011de}. Moreover, the experimental impact of the photon-dark-photon kinetic mixing has received tremendous interests. Currently, an industry has formed utilizing the existing or building novel experiments to look for light and weakly-coupled new particles~\cite{Essig:2013lka, Alexander:2016aln}.

The goal of our work is the following. We will explore the possibility of having sizable dark matter self-interaction and at the same time having a sizable photon-dark-photon kinetic mixing so as to make the dark sector accessible to the laboratories. In other words, we want to ask the questions how visible a SIDM candidate can be, and if visible enough what are the leading experimental channels for probing it.

Before proceeding, we find a couple of clarifications of our motivation necessary. First, although SIDM is quite an attractive scenario offering special correlations between the dark matter and dark photon masses (especially when many other dark matter anomalies have faded away these days), there are still ongoing debates about to what extent the self-interaction is needed as the solution to the small-scale structure problems. See, {\it e.g.},~\cite{hopkins}, which simply resorts to baryonic physics. The fate of SIDM will eventually be dictated by more precise observations and simulations of dwarf galaxy formation. In this work, we would like to keep an open mind and consider a very wide range of dark matter self-interaction cross sections, instead of just the most preferred value for SIDM. We expect our results to be more useful this way.
Second, one might argue that the laboratory detection of SIDM is not guaranteed, because setting the photon-dark-photon kinetic mixing parameter to zero does not upset the SIDM picture which involves only the dark sector interactions. There is nothing wrong with this argument. However, if it were the case, the dark sector would be totally decoupled from the standard model sector, and the only hope to further explore the nature of dark matter would be through astrophysical observations (see, {\it e.g.},~\cite{Agrawal:2016quu}). This is not the direction we would like to go here.

For a pure-Dirac fermion dark matter, shortly after the parameter space of SIDM was obtained, it was pointed out that if the dark matter mass lies above a few GeV, direct detection experiments place by far the strongest upper bound on the kinetic mixing parameter~\cite{Kaplinghat:2013yxa, Zhang:2015era}, largely due to the small dark photon mass ($\lesssim 100\,$MeV) required by SIDM. For a (sub-)GeV scale SIDM, the direct detection limits are weaker and the low-energy high-intensity searches for dark photon and/or dark matter become more important. In this region, it has been pointed out that with a sufficiently large dark gauge coupling, the $B$-factories play the leading role in probing the SIDM~\cite{An:2015pva}.

On the other hand, because the dark photon is already massive in the model, it is allowed to introduce a Majorana mass term to the dark matter. If the Majorana mass is much smaller than the Dirac mass, the dark matter becomes pseudo-Dirac, {\it i.e.}, the physical states are two Majorana particles with a small mass splitting of order the Majorana mass. Assuming dark matter particles are all in the lighter state, the scattering of dark matter on nucleus target via a dark photon exchange at tree level has to convert it to the heavier state. If this mass splitting is much larger than the typical incoming kinetic energy in the center-of-mass frame, the up-scattering process simply cannot happen and the strong direct detection constraints are evaded. At the same time, one has to worry what will happen to the SIDM in the presence of such a mass splitting. Naively, the dark matter self scattering with a tree-level dark photon exchange will also become up scattering from two lighter states to two heavier ones. If the kinetic energy is not large enough to compensate for the splitting, one may have to go to the next order and consider elastic scattering with two dark photon exchange (a box diagram, see below). If this were the case, the expected self-interaction cross section would be much smaller than that for a pure-Dirac dark matter.

As the main result of this work, our calculation reveals another key quantity, $\alpha_D^2 m_D$, where $\alpha_D$ is the dark fine-structure constant and $m_D$ is the dark matter mass. Unlike the direct detection process, the dark matter particle could gain a potential energy as large as $\alpha_D^2 m_D$ during the low velocity ($v\ll \alpha_D$) self-scattering process. We find that if the available potential energy is large enough to compensate for the mass splitting, the up-scattering process becomes kinematically allowed within the potential well. In this case, the self-interaction cross section for a pseudo-Dirac fermion dark matter is not suppressed and remains comparable to the pure-Dirac fermion case. If we further increase the mass splitting beyond $\alpha_D^2 m_D$, the quantum mechanical effect stops being effective. As a result, the up-scattering is forbidden everywhere and the dark matter self-interaction potential becomes genuinely loop suppressed. To maintain as large self scattering cross section, one must resort to much smaller dark photon mass. Based on these observations, we derive the new parameter space for a pseudo-Dirac SIDM.

This paper is organized as follows. In section II we introduce the massive dark QED model and discuss the role of adding a small Majorana mass to the dark matter in avoiding the strong direct detection constraints. In sections III, IV, V, we describe our method of calculating the dark matter elastic self-scattering cross section at low velocities, taking into account of the non-perturbative effects. Our numerical results are presented in section VI. We will conclude and outline several possible channels for probing the pseudo-Dirac SIDM in the future, in particular at high energy and intensity collider experiments.

We note that the dark matter self-interaction in the presence of mass splitting has been explored in~\cite{Schutz:2014nka}, but their discussion focused on very small mass splitting less than $\sim10$\,keV. In that case the inelastic scattering of dark matter in direct detection is still kinematically allowed and the constraints on the kinetic mixing parameter remain very strong. The impact of the mass splitting $\Delta m$ is less significant than what we shall show below.

\section{Model}

The Lagrangian for the massive dark QED model is
\begin{eqnarray}\label{Lfull}
\mathcal{L} &=& \mathcal{L}_{\rm SM} + \bar\chi i\gamma^\mu(\partial_\mu - i g_D V_\mu) \chi - m_D \bar \chi \chi - \frac{\Delta m}{4} \bar \chi^c \chi - \frac{\Delta m}{4} \bar \chi \chi^c -\frac{1}{4} V_{\mu\nu} V^{\mu\nu} + \frac{1}{2} m^2_{V} V_\mu V^\mu - \frac{\kappa}{2}F_{\mu\nu} V^{\mu\nu} \ ,
\end{eqnarray}
where in the limit $\Delta m=0$ the Dirac fermion $\chi$ is the dark matter field, $V_\mu$ is the dark photon, $g_D$ is the dark gauge coupling ($\alpha_D=g_D^2/(4\pi)$ will be the dark fine-structure constant) and $\kappa$ is the kinetic mixing between the photon and the vector field $V$. The $U(1)$ kinetic mixing term can be removed at the price of redefining the photon field $A_\mu \to A_\mu + \kappa V_\mu$. This will result in an effective coupling between the dark photon and the usual electromagnetic current (made of standard model particles), $\kappa e J_{\rm em}^\mu V_\mu$.

\begin{figure}[h]
\centerline{\includegraphics[width=8cm]{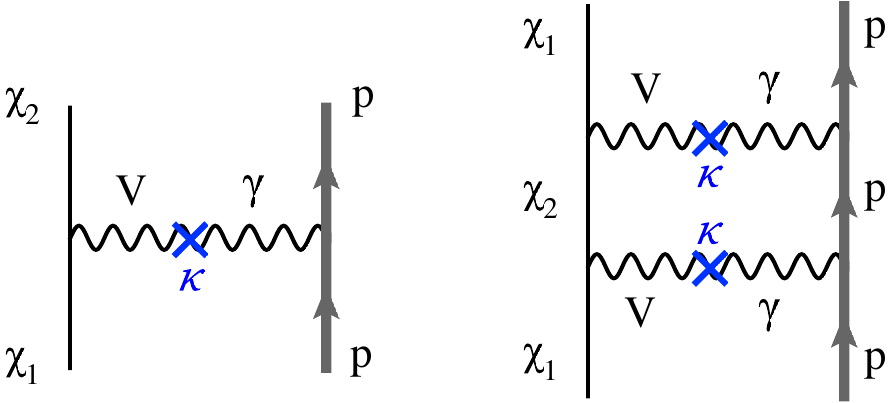}}
\caption{Diagrams for dark matter direct detection. The loop diagram with crossed dark photon propagators is not shown.}\label{-1}
\end{figure}

Turning on the Majorana mass $\Delta m$ will split $\chi$ into two Majorana fermion mass eigenstates,
\begin{eqnarray}\label{chi12}
\chi_{1} = \frac{i}{\sqrt2}(\chi - \chi^c), \hspace{0.5cm}\chi_{2} = \frac{1}{\sqrt2}(\chi + \chi^c), \hspace{0.5cm} m_{1,2} = m_D \mp \frac{1}{2}\Delta m \ .
\end{eqnarray}
where $\chi^c$ is the charge-conjugation of $\chi$ field. Throughout the paper we assume that all the dark matter today are in the lighter state $\chi_1$.
We also assume that $\Delta m$ is real and much smaller than the Dirac mass $m_D$.
The dark gauge interaction vertex becomes off-diagonal with respect to $\chi_1$ and $\chi_2$
\begin{eqnarray}
\mathcal{L}_{\rm int} = \frac{i}{2} g_D \bar \chi_2 \gamma^\mu \chi_1 V_\mu + {\rm h.c.} \ .
\end{eqnarray}
As a result, the tree level scattering of dark matter $\chi_1$ on the proton target will convert it into the heavier state $\chi_2$ (Fig.~\ref{-1}). Near the earth, the dark matter velocity distribution is peaked at $\sim 10^{-3} c$. Therefore, for most targets the typical kinetic energy in the dark matter-nucleus system is at most a few hundred keV. If the $\chi_1-\chi_2$ mass difference $\Delta m$ is greater than an MeV, the up-scattering process $\chi_1 p \to \chi_2 p$ is kinematically forbidden and the tree-level direct detection constraint is simply evaded. At next order, one could consider elastic scattering $\chi_1 p \to \chi_1 p$ happening at loop level with two dark photon exchange (see right plot of Fig.~\ref{-1}), but that will cost an additional power of $(\kappa \alpha \alpha_D)$ as well as the loop factor in the amplitude. Given the existing upper limit on $\kappa$ (see Fig.~4 in Ref.~\cite{Alexander:2016aln}), such a contribution cannot lead to a competitive constraint.

Adding a Majorana dark matter mass will also generate a contribution to the dark photon mass at loop level, which is of order $\delta m_V^2 \sim \frac{\alpha_D}{4\pi} |\Delta m|^2 \ln({\Lambda_{\rm UV}}/{m_D})$, where $\Lambda_{\rm UV}$ is the cutoff scale where the dark $U(1)$ is broken. This will not bring about unnaturalness as long as the dark photon bare mass satisfies $m_V \gtrsim \delta m_V$.

\section{Dark Matter Self-scattering Diagrams}\label{sec3}

\begin{figure}[h]
\centerline{\includegraphics[width=17cm]{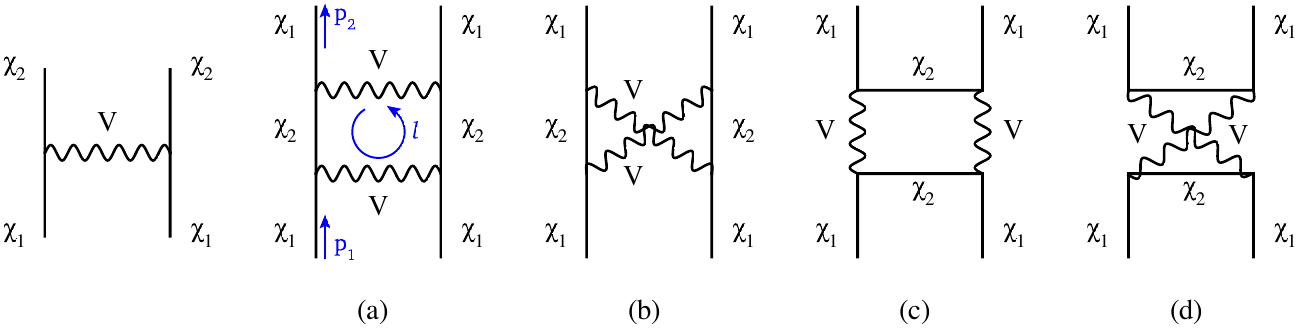}}
\caption{Tree-level and one-loop diagrams for dark matter self-scattering process, through one or two dark photon exchange.}\label{0}
\end{figure}

The main goal of this work is to calculate the cross section of dark matter self-interaction for a pseudo-Dirac dark matter. Because the dark matter velocity is much lower in dwarf galaxies ($\sim10^{-4}$) than in the local dark matter wind ($\sim 10^{-3}$), naively if the above mass splitting $\Delta m$ is large enough to forbid the up scattering in direct detection, the tree level up-scattering processes $\chi_1\chi_1\to \chi_2\chi_2$ (see the left plot in Fig.~\ref{0}) will also be kinematically forbidden. 
The next step seems to be considering $\chi_1\chi_1\to\chi_1\chi_1$ scattering at loop level by integrating out $\chi_2$, also shown in Fig.~\ref{0}. 
However, this argument is only valid for very large mass splitting $\Delta m$. In the $\Delta m\to0$ limit (pure Dirac dark matter case), the loop diagrams by themselves are not able to reproduce the leading Yukawa potential, $-\alpha_D e^{-m_V r}/r$, but rather the $\alpha_D^2$ order correction to it (for such a calculation in real QED, see~\cite{Schwinger:1949zz}). When $\Delta m$ is small enough for quantum mechanical effects to be important, we must keep the $\chi_2$ state in the calculation. 

Because we consider non-relativistic dark matter particle scatterings, it would greatly simplify the calculation by first going to the non-relativistic effective theory of the massive dark QED. According to the power counting in NRQED~\cite{Caswell:1985ui, Kinoshita:1995mt}, the leading contributions are made from Coulomb dark photon propagating in the loop. The radiative dark photon contributions are suppressed by the dark matter velocity at each vertex. Including only Coulomb dark photons, it is easiest to find that the diagrams (b) contribution vanishes because in the $l^0$ integral ($l$ is the loop momentum), both poles of the integrand are located on the same side of the real axis. 
Also, because the Coulomb photon exchange must be instantaneous, the diagrams (c,\,d) will not contribute at leading order.
Only diagram (a) survives and the dominant contribution to the scattering amplitude is when the loop momentum $l$ is of same order as the total momentum transfer $q=p_2-p_1$. This corresponds to each of the dark photon propagator only carrying a momentum much smaller than the dark matter mass scale $m_1\sim m_D$ and the two $\chi_2$ propagators are only off-shell by the amount of the momentum transfer $q$ or the mass difference $\Delta m$. The low loop momentum dominance nature of the diagram justifies the use of Schr\"odiger equation to resum the multiple Coulomb dark photon exchange.

\section{Non-relativistic Hamiltonian}

As discussed above, we make non-relativistic expansion of the dark matter field, with the definition
\begin{eqnarray}
\chi = \sqrt{\frac{E+m_D}{2E}} \left( \begin{array}{c}
e^{-i m_D t} \xi + i e^{i m_D t} \frac{\vec \sigma \cdot \vec \nabla}{E+m_D} \eta \\
e^{i m_D t} \eta - i e^{-i m_D t} \frac{\vec \sigma \cdot \vec \nabla}{E+m_D} \xi
\end{array} \right) \ .
\end{eqnarray}
We define $\chi_{(+)}=\xi$ and $\chi_{(-)}=i\sigma_2 \eta^*$. In the limit $\Delta m=0$, $\chi_{(+)}$ annihilates a Dirac dark matter particle while $\chi_{(-)}$ annihilates an anti-dark matter particle. Together with the full theory Lagrangian (\ref{Lfull}), we obtain the non-relativistic Hamiltonian at lading order, 
\begin{eqnarray}
H_{\rm dark-NRQED} = \chi_{(+)}^\dagger \left[ - \frac{\nabla^2}{2 m_D} + g_D V_0 \right] \chi_{(+)} +  \chi_{(-)}^\dagger \left[ - \frac{\nabla^2}{2 m_D} - g_D V_0 \right] \chi_{(-)} + \frac{\Delta m}{2}  \left[ \chi_{(+)}^\dagger  \chi_{(-)} + \chi_{(-)}^\dagger  \chi_{(+)} \right] \ .
\end{eqnarray}
Next, we go to the basis where $H_{\rm dark-NRQED}$ is diagonal for free dark matter particles at infinity where $V_0=0$, with the rotation
\begin{eqnarray}
\left(\begin{array}{c}
\chi_{(+)} \\
\chi_{(-)}
\end{array} \right) = 
\frac{1}{\sqrt2}\left(\begin{array}{cc}
1 & 1 \\
-1 & 1
\end{array} \right) \left(\begin{array}{c}
\chi_{1} \\
\chi_{2}
\end{array} \right) \ .
\end{eqnarray}
Hereafter, $\chi_{1,2}$ denote the non-relativistic counterpart of those fields defined in Eq.~(\ref{chi12}).
In the basis of $\{\chi_1, \chi_2\}$, the single-particle non-relativistic Hamiltonian takes the form
\begin{eqnarray}
H_{\rm dark\ NRQED} =  \left(
\chi_{1}^\dagger \ 
\chi_{2}^\dagger \right) \left(\begin{array}{cc}
- \frac{\nabla^2}{2 m_D} & g_D V_0 \\
g_D V_0 & - \frac{\nabla^2}{2 m_D} + \Delta m
\end{array} \right)  \left(\begin{array}{c}
\chi_{1} \\
\chi_{2}
\end{array} \right) \ ,
\end{eqnarray}
where we shifted $H_{\rm dark\ NRQED}$ by a term proportional to the unit matrix such that the lighter state $\chi_1$ (the dark matter) has only kinetic energy when $V_0=0$. This does not affect the evolution of states or our result.

Because we will be discussing the $\chi_1\chi_1$ scattering process with $\chi_2\chi_2$ as a possible intermediate state, it is convenient to derive the two-body state effective potential~\cite{Hisano:2004ds}. In the basis of $\{\Psi_1=\chi_1\chi_1, \Psi_2=\chi_2\chi_2\}$, it takes the form
\begin{eqnarray}\label{Veff}
V(r) = \left(\begin{array}{cc}
0 & \frac{\alpha_D}{r}e^{- m_V r} \\
\frac{\alpha_D}{r}e^{- m_V r} & 2\Delta m
\end{array} \right) \ .
\end{eqnarray}
We expand the continuum state wavefunction as
\begin{eqnarray}
\Psi_{i=1,2}(\vec r) = \sum_{\ell m} R^{(i)}_{k\ell}(r) Y_{\ell m}(\hat r) Y_{\ell m}^*(\hat k) \ ,
\end{eqnarray}
where $\vec k$ is the relative momentum of two dark matter particles at infinity. Then the Schr\"odinger equation for each partial wave is
\begin{eqnarray}\label{SE}
\frac{1}{2\mu}\left(\frac{1}{r^2}\frac{d}{dr} \left( r^2 \frac{d}{dr} \right)  - \frac{\ell(\ell+1)}{r^2}\right)
\left(\begin{array}{c}
R_{k\ell}^{(1)} \\
R_{k\ell}^{(2)}
\end{array} \right) + \left(\begin{array}{cc}
\frac{k^2}{2\mu} & \frac{\alpha_D}{r}e^{- m_V r} \\
\frac{\alpha_D}{r}e^{- m_V r} & \frac{k^2}{2\mu} + 2 \Delta m
\end{array} \right)
\left(\begin{array}{c}
R_{k\ell}^{(1)} \\
R_{k\ell}^{(2)}
\end{array} \right) = 0\ ,
\end{eqnarray}
where $\mu \simeq m_D/2$ is their reduced mass of two dark matter particle system. In the parameter space of interest to this work $\Delta m \gg k^2/(2\mu)$, we need to solve the Schr\"odinger equation with the boundary condition at infinity,
\begin{eqnarray}\label{BC}
R_{k\ell}^{(1)} (r\to \infty) = \frac{4\pi}{kr} i^\ell e^{i\delta_\ell} \cos\left( kr - (\ell+1)\frac{\pi}{2} + \delta_\ell \right) \ , \hspace{0.5cm} R_{k\ell}^{(2)} (r\to \infty) = 0 \ ,
\end{eqnarray}
where $\delta_\ell$ is the phase shift for each partial wave $\ell$.

The momentum-transfer scattering cross section between two dark matter particles, commonly used for exploring the solution to small-scale problems, is defined as~\cite{Krstic1999}
\begin{eqnarray}\label{cross}
\sigma_T = \int d\Omega (1-\cos\theta) \frac{d \sigma}{d\Omega} = \frac{4\pi}{k^2} \sum_{\ell=0}^{\infty} (l+1) \sin^2(\delta_{\ell+1} - \delta_\ell) \ .
\end{eqnarray}

\section{Adiabatic Approximation}

In principle, we need to solve the scattering problem using the $2\times2$ coupled equations (Eq.~(\ref{SE})). We find that for the parameter space of interest to the present work, the adiabatic condition is valid which could greatly simplify Eq.~(\ref{SE}) back to one Schr\"odinger equation.

\begin{figure}[t]
\centerline{\includegraphics[width=7cm]{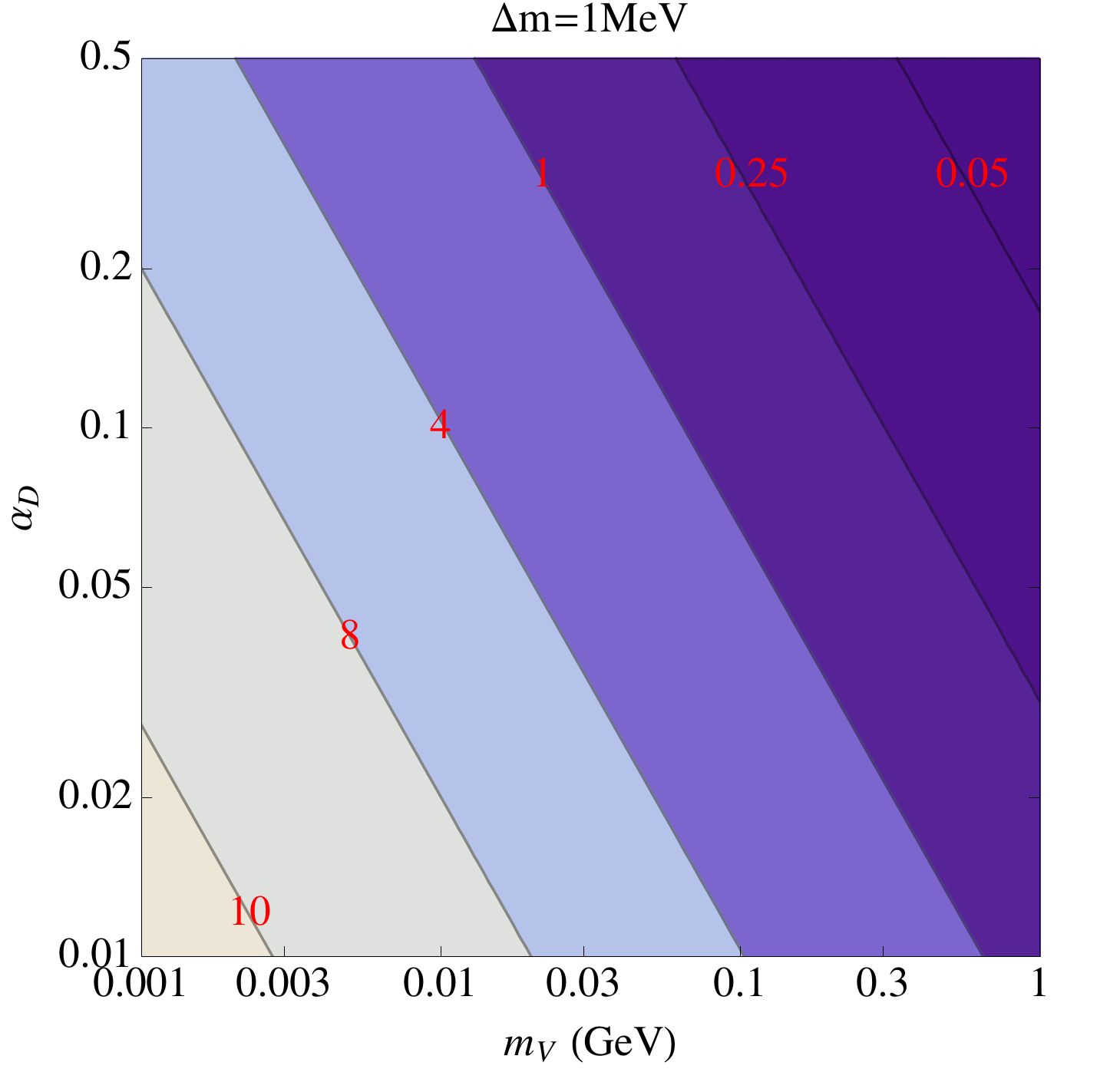} \includegraphics[width=7cm]{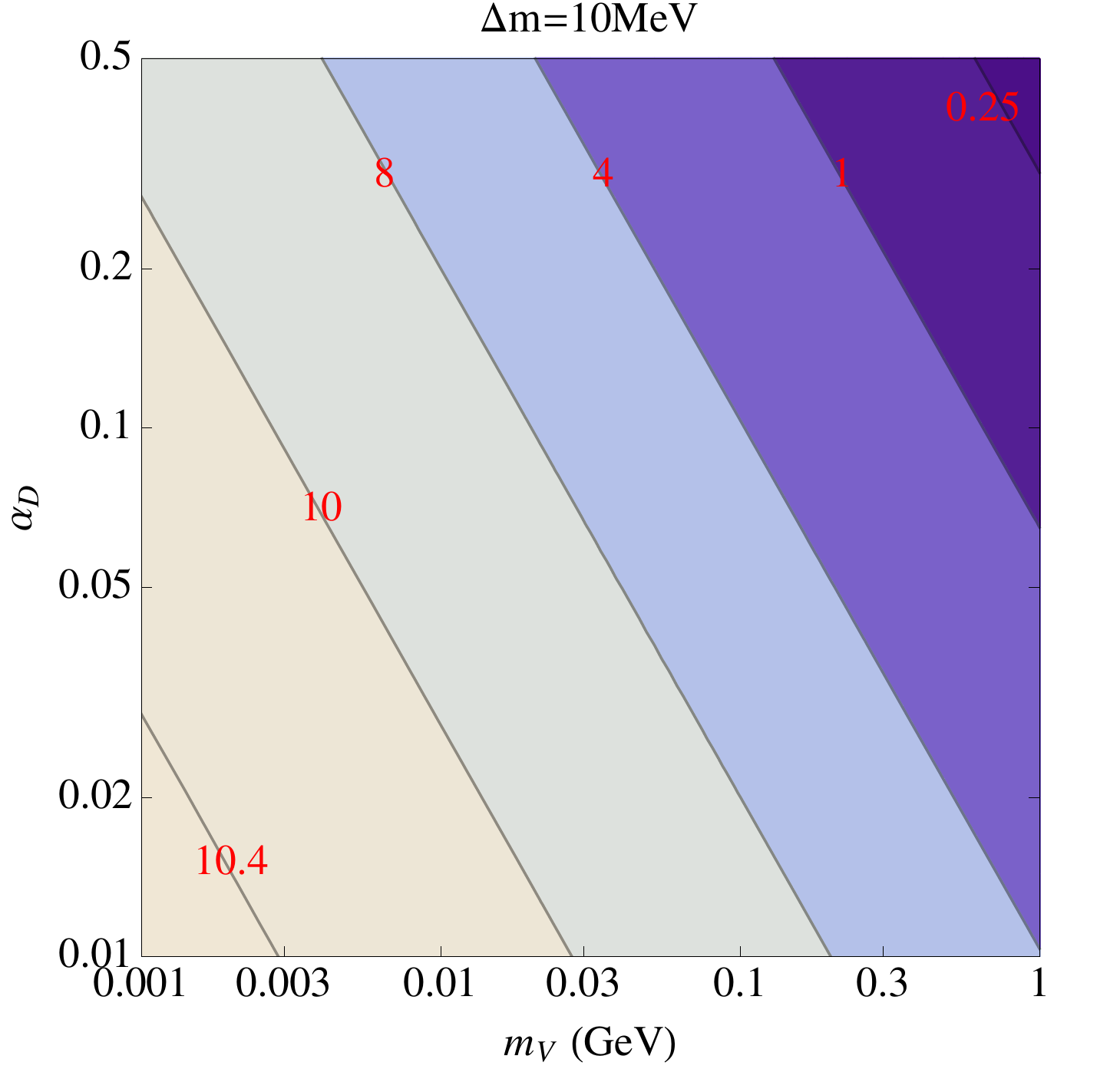}}
\caption{Contours of constant ratio $\Delta E(r)/|\dot\Theta(r)|$ in the $m_V-\alpha_D$ parameter space for two choices of $\Delta m$ values.}\label{1}
\end{figure}

It is useful to define $R_{k\ell}^{(i)} = r^{\ell-1} \phi_{k\ell}^{(i)}$ in Eq.~(\ref{SE}). As such, we conveniently have boundary conditions at the origin, $\phi_{k\ell}^{(i)}(0)=0$, and $\frac{d}{dr}\phi_{k\ell}^{(1,2)}(0)$ are the two parameters one needs to adjust in order to satisfy to the boundary conditions at infinity, Eq.~(\ref{BC}).
The equation for $\phi_{k\ell}^{(i)}$ takes the form
\begin{eqnarray}\label{SE2}
\frac{1}{2\mu}\left(\frac{d^2}{dr^2} + \frac{2\ell}{r} \frac{d}{dr} - \frac{2\ell}{r^2}\right)
\left(\begin{array}{c}
\phi_{k\ell}^{(1)} \\
\phi_{k\ell}^{(2)}
\end{array} \right) + \left(\begin{array}{cc}
\frac{k^2}{2\mu} & \frac{\alpha_D}{r}e^{- m_V r} \\
\frac{\alpha_D}{r}e^{- m_V r} & \frac{k^2}{2\mu} + 2 \Delta m
\end{array} \right)
\left(\begin{array}{c}
\phi_{k\ell}^{(1)} \\
\phi_{k\ell}^{(2)}
\end{array} \right) = 0 \ .
\end{eqnarray}
In order to proceed, we find the adiabatic approximation very useful here. At every $r$, we can use the rotation angle $\Theta(r)$ to diagonalize the potential $V$ in Eq.~(\ref{Veff}), and obtain the energy splitting $\Delta E(r)$,
\begin{eqnarray}
\Theta(r) = - \frac{1}{2} \arctan\left(\frac{\alpha_D e^{-m_V r}}{\Delta m \ r}\right) \ , \hspace{0.5cm}
\Delta E(r) = 2 \sqrt{\Delta m^2 + \left(\frac{\alpha_D e^{-m_V r}}{r}\right)^2} \ .
\end{eqnarray}
The adiabatic condition~\cite{landau, Mikheev:1986gs} is fulfilled if the time derivative $|\dot\Theta(r)|$ is much smaller than $\Delta E(r)$, where we also work under the approximation that $|\dot\Theta(r)| \lesssim \alpha_D |d\Theta(r)/dr|$, {\it i.e.}, the dark matter particle can be accelerated up to $dr/dt\sim\alpha_D$ at most in the potential well. Clearly, the larger the mass splitting $\Delta m$ and the smaller the dark coupling $\alpha_D$, the easier it is to satisfy the adiabatic condition. In Fig.~\ref{1}, we plot the ratio $\Delta E(r)/|\dot\Theta(r)|$ in the $m_V-\alpha_D$ parameter space for $\Delta m=1$\,MeV and $10\,$MeV, respectively.
In the following numerical calculations we will take a benchmark value $\alpha_D=10^{-2}$. In this case, the condition $\Delta E(r)/|\dot\Theta(r)|>1$ is well satisfied for most of the self-interacting dark matter parameter space in our results.
In contrast, the mass splitting considered in~\cite{Schutz:2014nka} is much smaller than here and the transition between the two energy levels cannot be neglected.

In the scattering problem, the dark matter state starts with $\phi_{k\ell}^{(1)}\neq 0$, $\phi_{k\ell}^{(2)}=0$ at infinity. As the two particles approach each other, the two energy levels begin to mix with each other when the off-diagonal potential terms turn on. Under the adiabatic approximation, we neglect the transition between $\phi_{k\ell}^{(1)}$ and $\phi_{k\ell}^{(2)}$. In other words, the dark matter state will continuously stay on $\phi_{k\ell}^{(1)}$ level and the phase shift out of the scattering process is generated simply because the energy eigenvalue value gets deformed by interaction. In this case, the Schr\"odinger equation for $\phi_{k\ell}^{(1)}$ is simplified to
\begin{eqnarray}\label{SE3}
\left[\frac{d^2}{dr^2} + \frac{2\ell}{r} \frac{d}{dr} - \frac{2\ell}{r^2} + k^2 - 2 \mu \Delta m + 2 \mu \sqrt{ \Delta m^2 + \left(\frac{\alpha_D e^{-m_V r}}{r}\right)^2 }\right]
\phi_{k\ell}^{(1)}(r)=0 \ .
\end{eqnarray}
Effectively, in this adiabatic approximation, we have integrated out the heavier two-particle state $\chi_2'(r)\chi_2'(r)$ at every $r$. It is crucial to note that $\chi_2'\chi_2'$ is not equivalent to the free particle state $\chi_2\chi_2$. In particular, deep inside the potential well if the potential energy is much larger than the mass splitting, $\chi_2'\chi_2'$ is a linear combination of $\chi_1\chi_1$ and $\chi_2\chi_2$ with rotation angle equal to $\pi/4$. This picture is in line with the argument in section~\ref{sec3} that $\chi_2$ state must be kept in the calculation, until we properly integrate out the instantaneous heavier state $\chi_2'\chi_2'$ (instead of $\chi_2\chi_2$). The resulting effective potential used for Schr\"odinger equation is
\begin{eqnarray}\label{Vadiabatic}
V_{\rm adiabatic}(r) = \Delta m - \sqrt{ \Delta m^2 + \left(\frac{\alpha_D e^{-m_V r}}{r}\right)^2 } \ .
\end{eqnarray}
It is worth noting that in the large mass splitting limit, the approximate form is $V_{\rm adiabatic}(r) \simeq \alpha_D^2e^{-2m_V r}/(2\Delta m\ r^2)$.

We numerically solve (\ref{SE3}) starting with the boundary condition $\phi_{k\ell}^{(1)}(0)=0$ and adjust $\frac{d}{dr}\phi_{k\ell}^{(1)}(0)$ so that at large $r$ the first boundary condition in (\ref{BC}) is satisfied. The phase shift is then obtained in the standard way at a matching point $r_m\gg 1/m_V$ using
\begin{eqnarray}
\delta_\ell = \frac{k r_m j'_\ell(k r_m) - \beta_\ell j_\ell(k r_m)}{k r_m n'_\ell(k r_m) - \beta_\ell n_\ell(k r_m)}, \hspace{0.5cm} \beta_\ell +1 = \left. \frac{\ell r^\ell \phi_{k\ell}^{(1)}(r) + r^{\ell+1} \phi_{k\ell}^{(1)\displaystyle{'}}(r) }{\phi_{k\ell}^{(1)}(r)} \right|_{r=r_m} \ .
\end{eqnarray}
In our calculation, we use $r_m = 10/m_V$. We take the dark matter relative velocity in dwarf galaxies to be $v=10^{-4}$, and fix the dark fine-structure constant $\alpha_D=10^{-2}$. Finally, for the sum over $\ell$ in (\ref{cross}), we truncate it when the value of $|\delta_\ell|$ consistently falls below 0.01 at large enough $\ell$.

\section{Results}

\begin{figure}[t]
\centerline{\includegraphics[width=10.5cm]{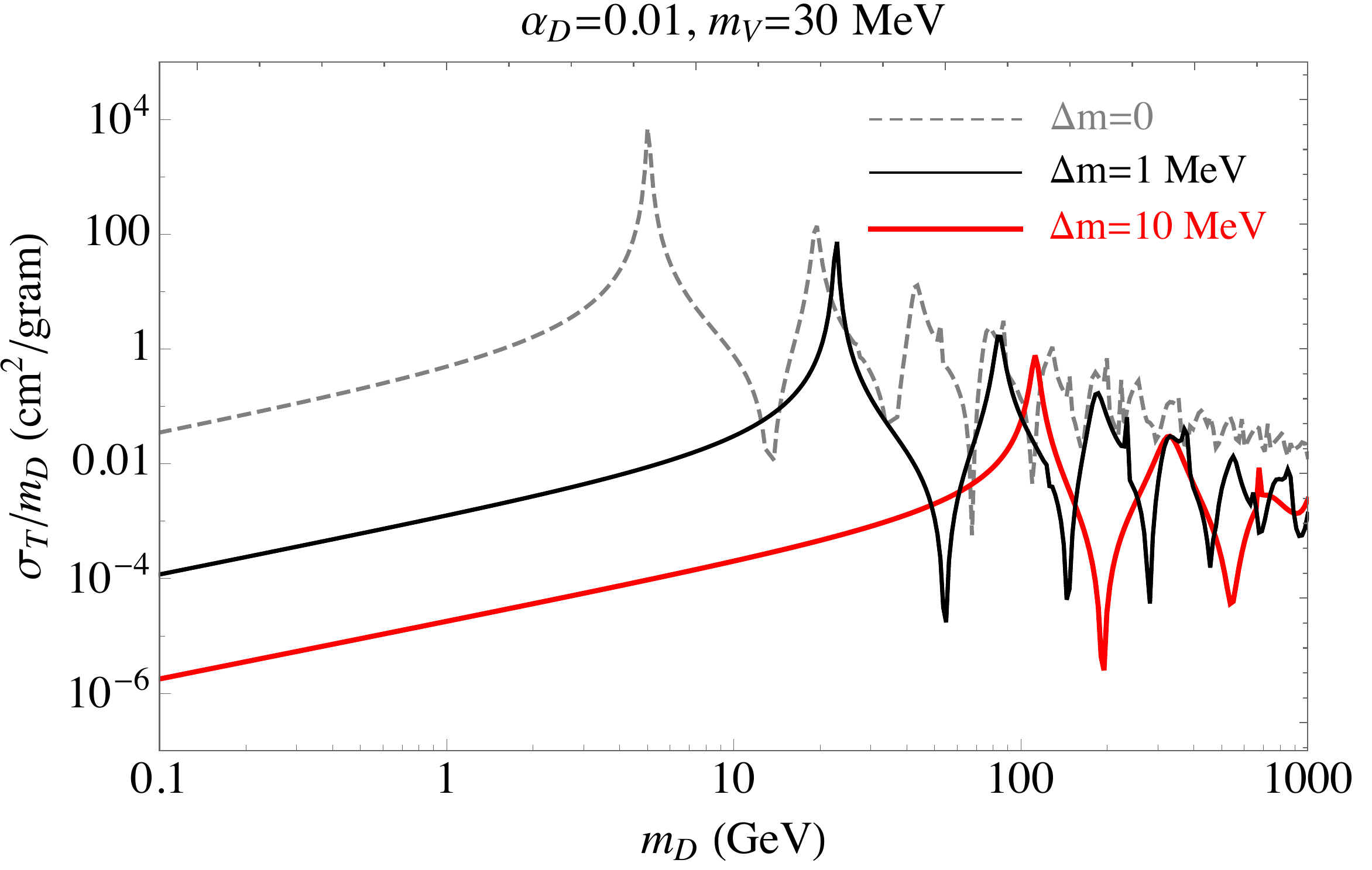}}
\caption{Ratio of the dark matter self-interaction cross section to its mass as a function of dark matter mass. We have fixed $\alpha_D=0.01$ and $m_V=30\,$MeV. The dashed (gray), solid (black) and thick (red) curves corresponds to pure-Dirac fermion dark matter and the pseudo-Dirac case with mass splitting equal to 0, 1, 10\,MeV, respectively.}\label{2}
\end{figure}

As motivated in the introduction, in this work we focus on the pseudo-Dirac dark matter case with a mass splitting large enough to suppress the direct detection constraint, $\Delta m > 1\,$MeV. Our new finding here is the importance of interplay between the two parameters, the mass splitting $\Delta m$, and the potential energy $\alpha_D^2 m_D$ (similar to the Bohr energy).

\begin{figure}[t]
\centerline{
\includegraphics[width=7.5cm]{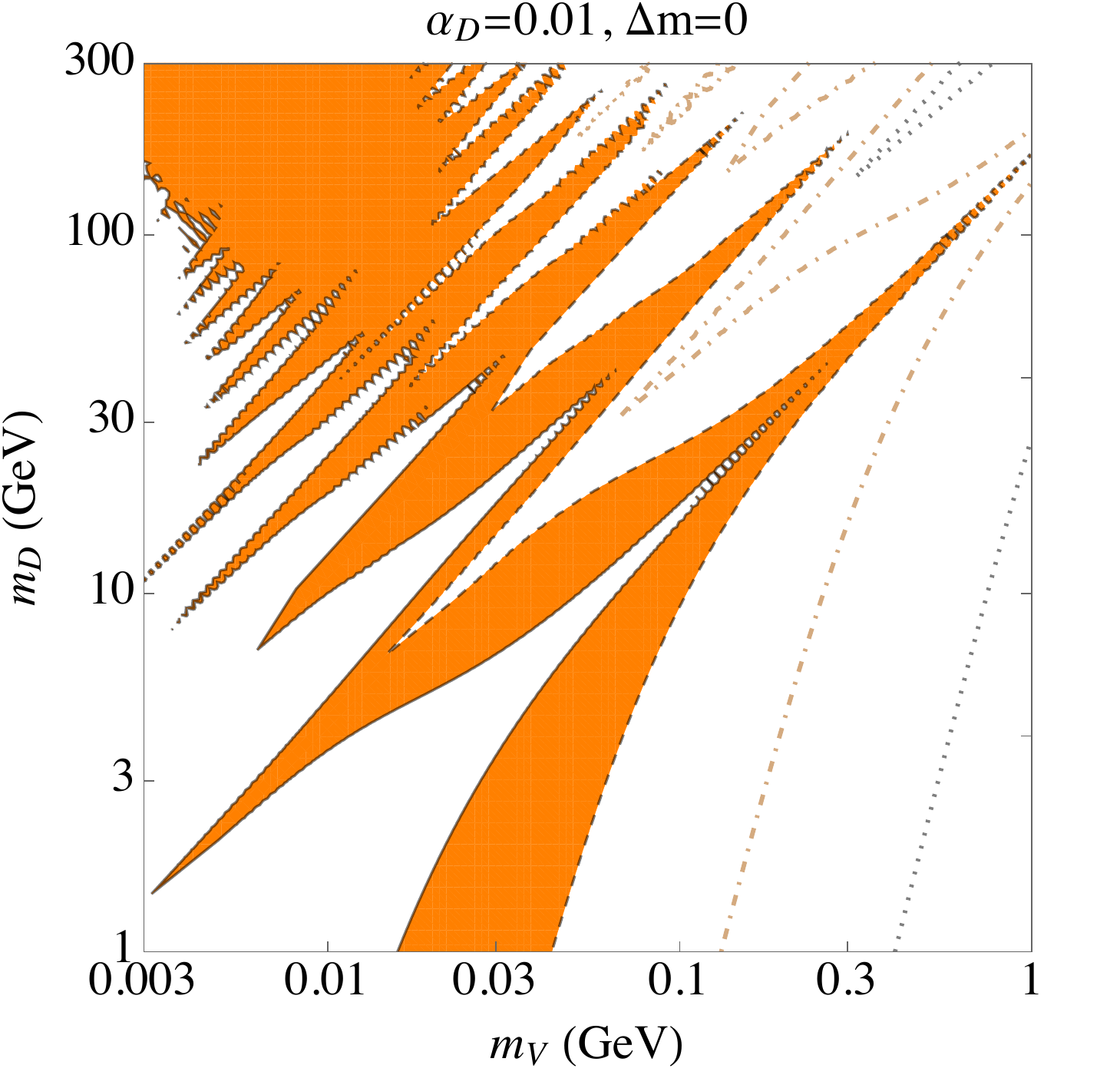}
\includegraphics[width=7.5cm]{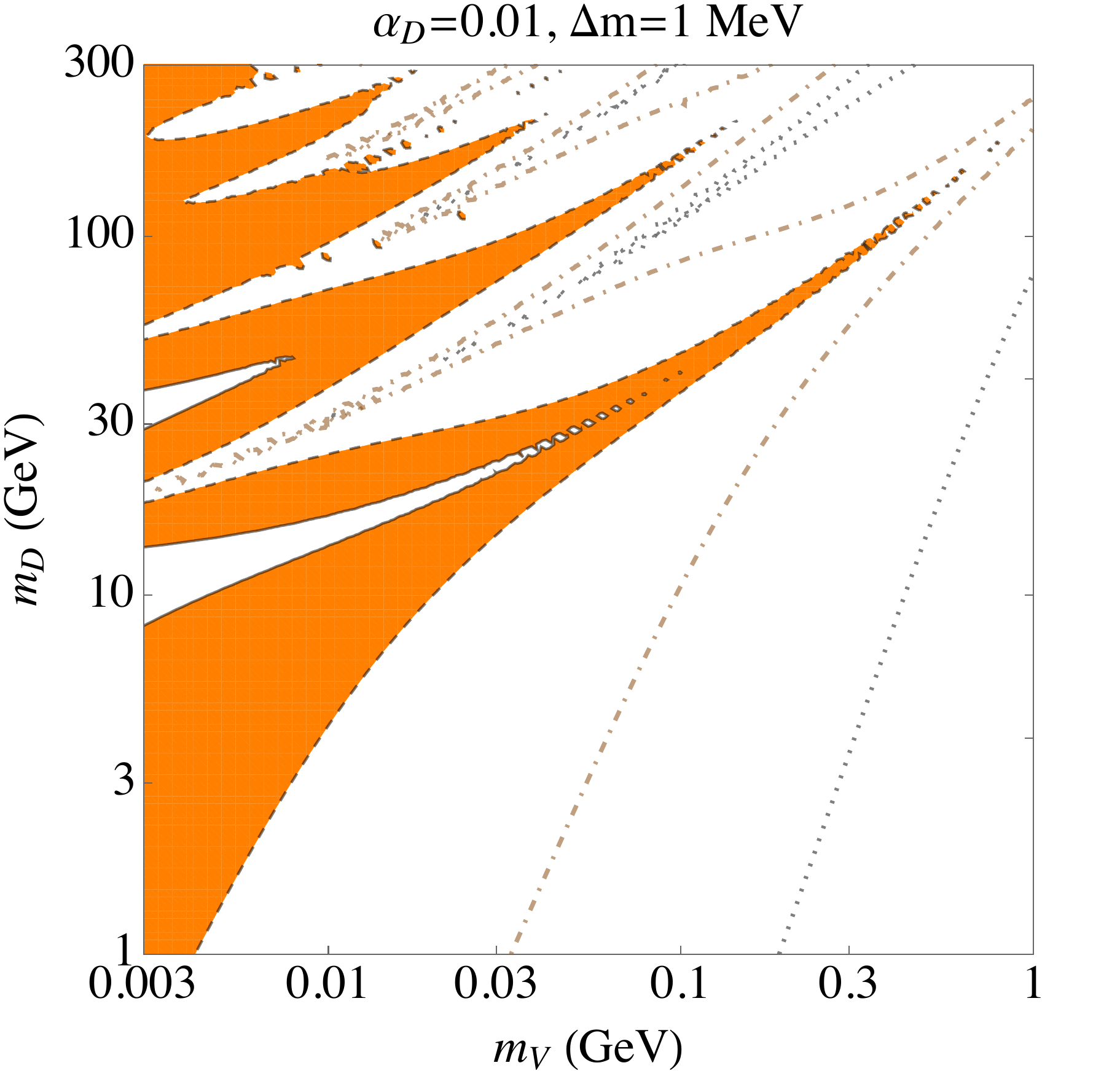}\vspace{3mm}}
\centerline{
\includegraphics[width=7.5cm]{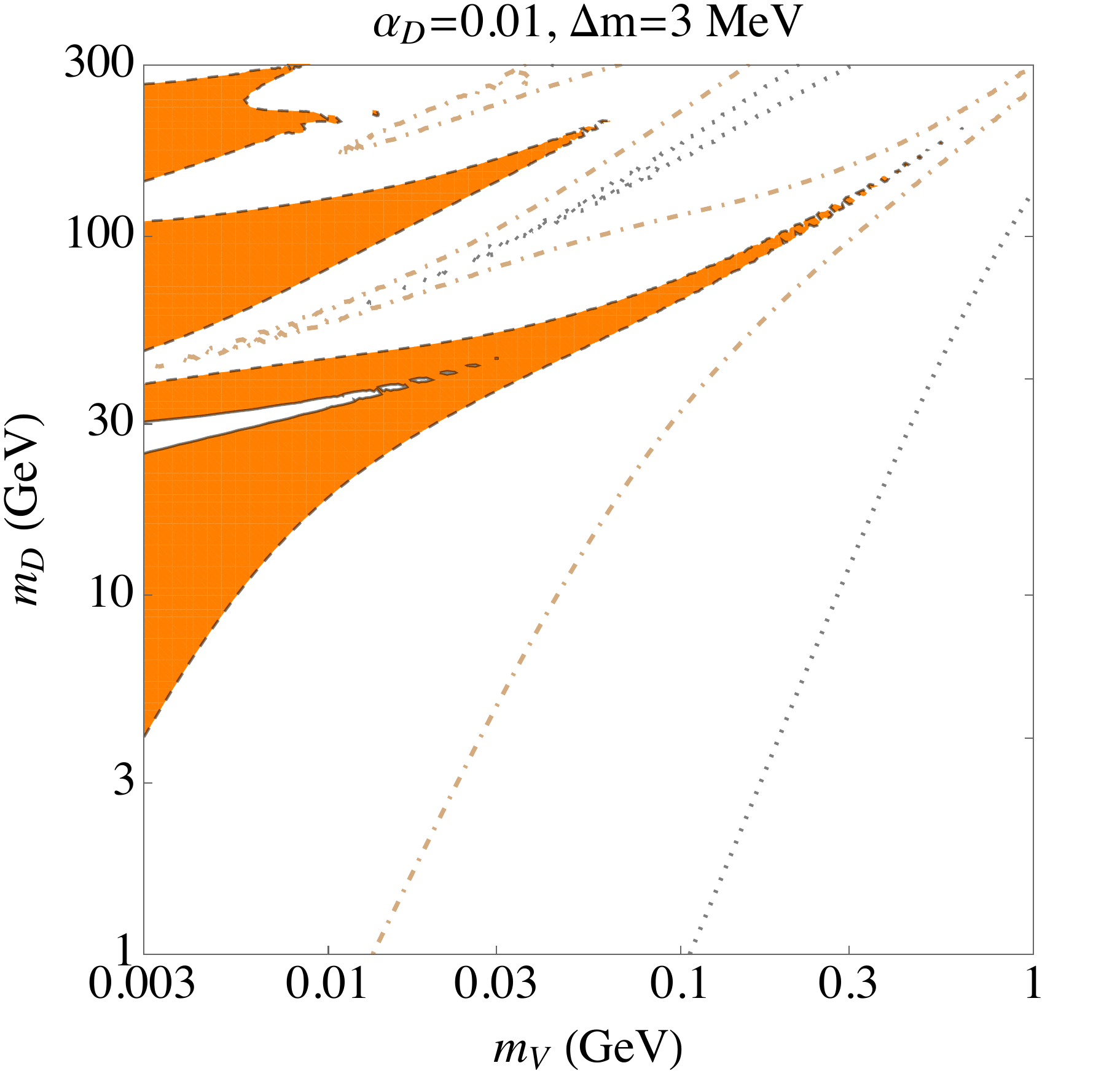}
\includegraphics[width=7.5cm]{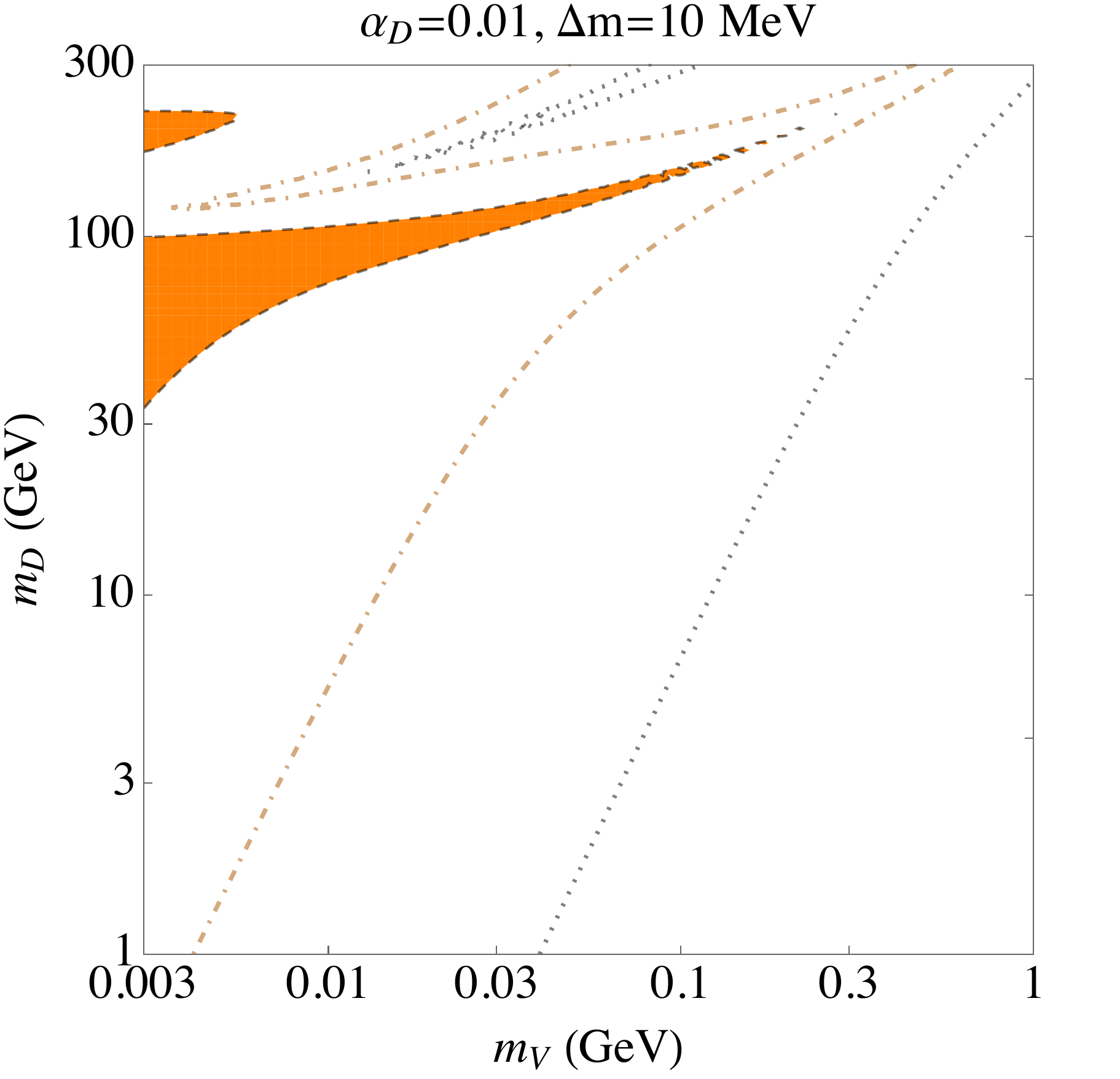}}
\caption{The orange region show the parameter space for a pure-Dirac fermion dark matter (upper-left) and pseudo-Dirac case with a mass splitting equal to 1\,MeV (upper-right), 3\,MeV (lower-left) and 10\,MeV (lower-right). In the orange region of each plot, the ratio of the dark matter self-interaction cross section to its mass $\sigma_T/m_D$ lies between $0.1\,{\rm cm^2/gram}$ (dahsed boundary) and 
$10\,{\rm cm^2/gram}$ (solid boundary). We also show contours with constant values $\sigma_T/m_D$ equal to $10^{-3}\,{\rm cm^2/gram}$ (orange dot-dahsed) and $10^{-5}\,{\rm cm^2/gram}$ (black dotted).
We do not show the region with even lower dark photon mass because the kinetic mixing parameter is very tightly constrained, $\kappa\lesssim 10^{-10}$ for $m_V\lesssim10\,$MeV~\cite{Essig:2013lka}, which conflicts with our goal of making the SIDM more visible.
}\label{3}
\end{figure}

If $\Delta m < \alpha_D^2 m_D$, the behavior of pseudo-Dirac dark matter self-interaction cross section has many similarities to the pure-Dirac case (for an anatomy of the latter, see~\cite{Tulin:2013teo}). For light enough dark matter when both $m_D \alpha_D$ and $m_D v$ are smaller than $m_V$, the self interaction can be correctly described by the Born approximation and the low energy $S$-wave scattering. As the dark matter mass grows such that $m_D v < m_V < m_D \alpha_D$, the interaction strength gets stronger, multiple dark photon exchanges become important and the scattering enters the quantum regime. In this regime, by increasing the dark photon mass $m_V$, the binding energies of $S$-wave bound states are modified. When one of the states becomes degenerate in energy with the initial state, the cross section $\sigma_T$ gets resonantly enhanced. 
At very large $m_D$, the dark photon mass become negligible compared to both $m_D \alpha_D$ and $m_D v$. 
In this case not only the quantum mechanics effects are important but also we have to sum up to a large number of partial waves $\ell$. 
Sometimes we could encounter the case where the resonant peaks due to higher-$\ell$ channels stand on top of an $S$-wave peak.
In Fig.~\ref{2}, we plot the ratio of dark matter self-interaction cross section over mass as a function of the dark matter mass, $m_D$, for three different values of $\Delta m$. In the right half of this plot, the condition $\Delta m < \alpha_D^2 m_D$ is satisfied for all the three curves. We find the position of the resonant peaks are different for pseudo-Dirac and pure-Dirac dark matter cases, and there are fewer of them in the former case, but the peaks are enhanced by similar amount. They key reason for the similar enhancement is that when two particles approach the inside of their potential well, the available potential energy can be large enough to compensate for the mass splitting and enables the $\chi_1\chi_1\to\chi_2\chi_2$ transition to still happen.

On the other hand, if $\alpha_D^2 m_D < \Delta m$, the mass splitting is so large that even the potential energy is not enough for $\chi_1$ to up-scatter into $\chi_2$. In this case, the leading contribution to the self-interaction comes from loop induced process $\chi_1\chi_1\to\chi_1\chi_1$ as discussed in section III (see also the discussion below Eq.~(\ref{Vadiabatic})), and the cross section gets much more suppressed. As we reduce the dark matter mass from right to left in Fig.~\ref{2}, when $\Delta m$ is large enough, it is possible to reach a situation where $\alpha_D^2 m_D < \Delta m$ and $\alpha_D m_D > m_V$ (for example, in the region $3\,$GeV$\,\lesssim m_D\lesssim100\,$GeV along the red curve). In this case, although the scattering is still in the quantum regime, the self interaction cross section already becomes loop suppressed and all the resonant peaks disappear.

In Fig.~\ref{3}, we plot, in the $m_D-m_V$ parameter space, contours of constant ratio of self-interaction cross section to mass of the dark matter, $\sigma_T/m_D$, for a pure-Dirac fermion dark matter, and for pseudo-Dirac fermion dark matter case with two mass splittings $\Delta m = 1,\, 3,\, 10\,$MeV. The orange region between the dashed and solid curves corresponds to $0.1\,{\rm cm^2/gram}\leq\sigma_T/m\leq10\,{\rm cm^2/gram}$~\cite{Spergel:1999mh}. 
Clearly, in the presence of a larger mass splitting, more SIDM regions shrink from the bottom of the plot.
In those regions that disappear, the mass splitting $\Delta m$ is larger than the potential energy $\alpha_D^2 m_D$.
As a result, the dark matter self interaction becomes loop suppressed and a large enough self-interaction cross section could be maintained only by resorting to a very light dark photon. This explains why the orange strips eventually disappears at the bottom-left corner of each plot. 

In this work, we would like to be open-minded by noting that, although SIDM is quite an attractive scenario offering special correlations between the dark matter and dark photon masses, there are still ongoing debates about to what extent the self-interaction is needed as the solution to the small-scale structure problems. 
To stay tuned for possible future variations of the favored value, we also show in Fig.~\ref{3} the contours with lower values of the self-interaction cross section, $\sigma_T/m_D=10^{-3}, \,10^{-5}\,{\rm cm^2/gram}$. In these cases, we find the dark photon is allowed to be much heavier in the light dark matter region (GeV-scale) and the interplay between $\Delta m$ and $\alpha_D^2 m_D$ as discussed above become less significant. This feature could also be read from Fig.~\ref{2}.

\begin{figure}[t]
\centerline{
\includegraphics[width=11cm]{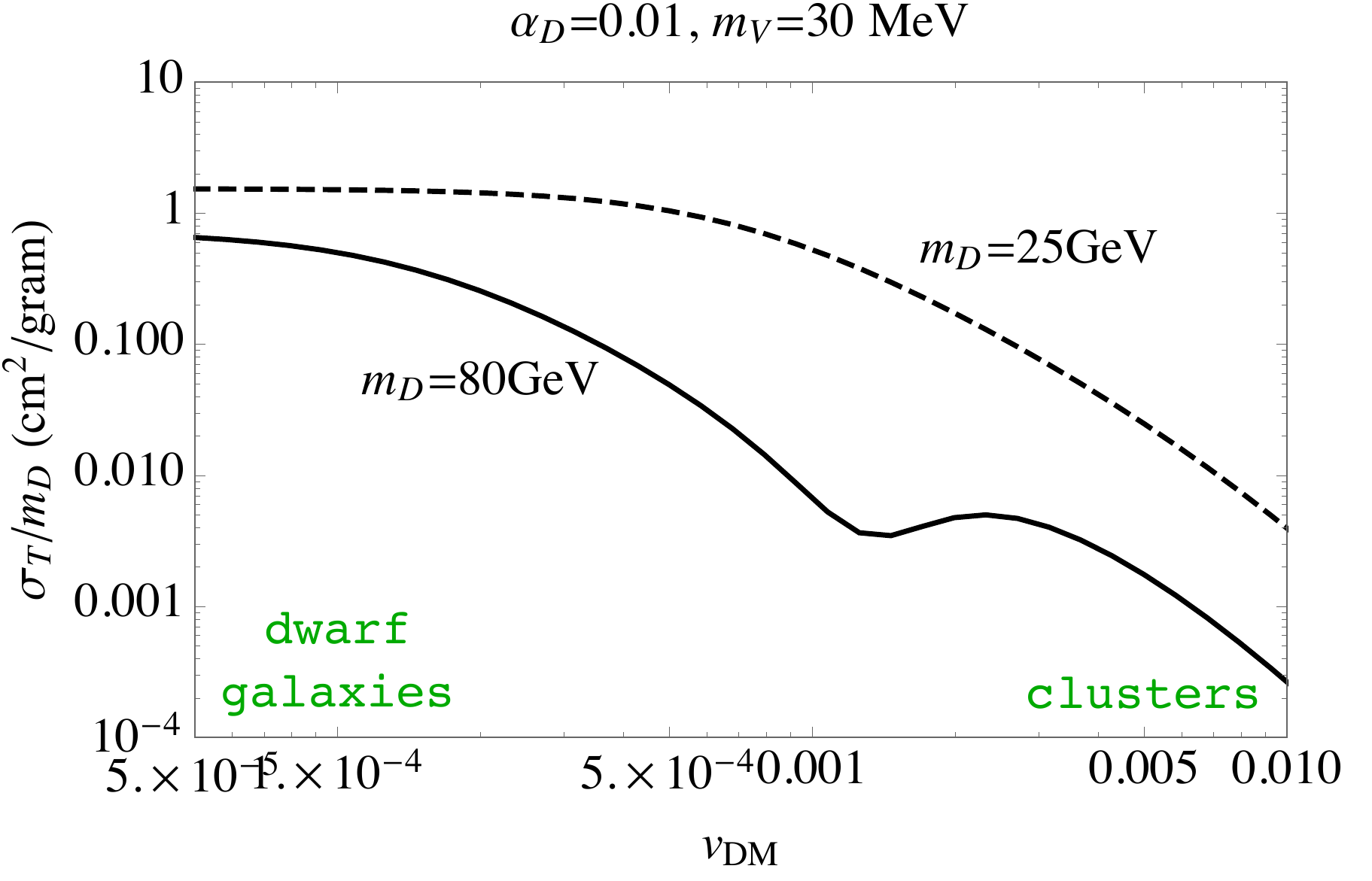}}
\caption{Velocity dependence in the dark matter self-scattering cross section, with $\Delta m=1\,$MeV.}\label{x}
\end{figure}

We find that with a larger mass splitting, the largest azimuthal quantum number $\ell$ needed to be summed up to gets smaller. The largest $\ell$ sum corresponds to the upper-left corners of each plot. We have also checked that, with MeV scale or larger mass splitting, the adiabatic approximation used in our calculation is valid in all regions relevant to the SIDM scenario. It is also crucial to check that the dark matter self-interaction is suppressed on the cluster scales, where the velocity is much higher ($v\sim 5\times10^{-3}c$), so that the constraints from bullet cluster observations~\cite{Randall:2007ph} are satisfied. Fig.~\ref{x} illustrates with two examples the point that in the presence of a Yukawa potential~\cite{Loeb:2010gj} this is indeed the case.

\begin{figure}[t]
\centerline{
\includegraphics[width=7.5cm]{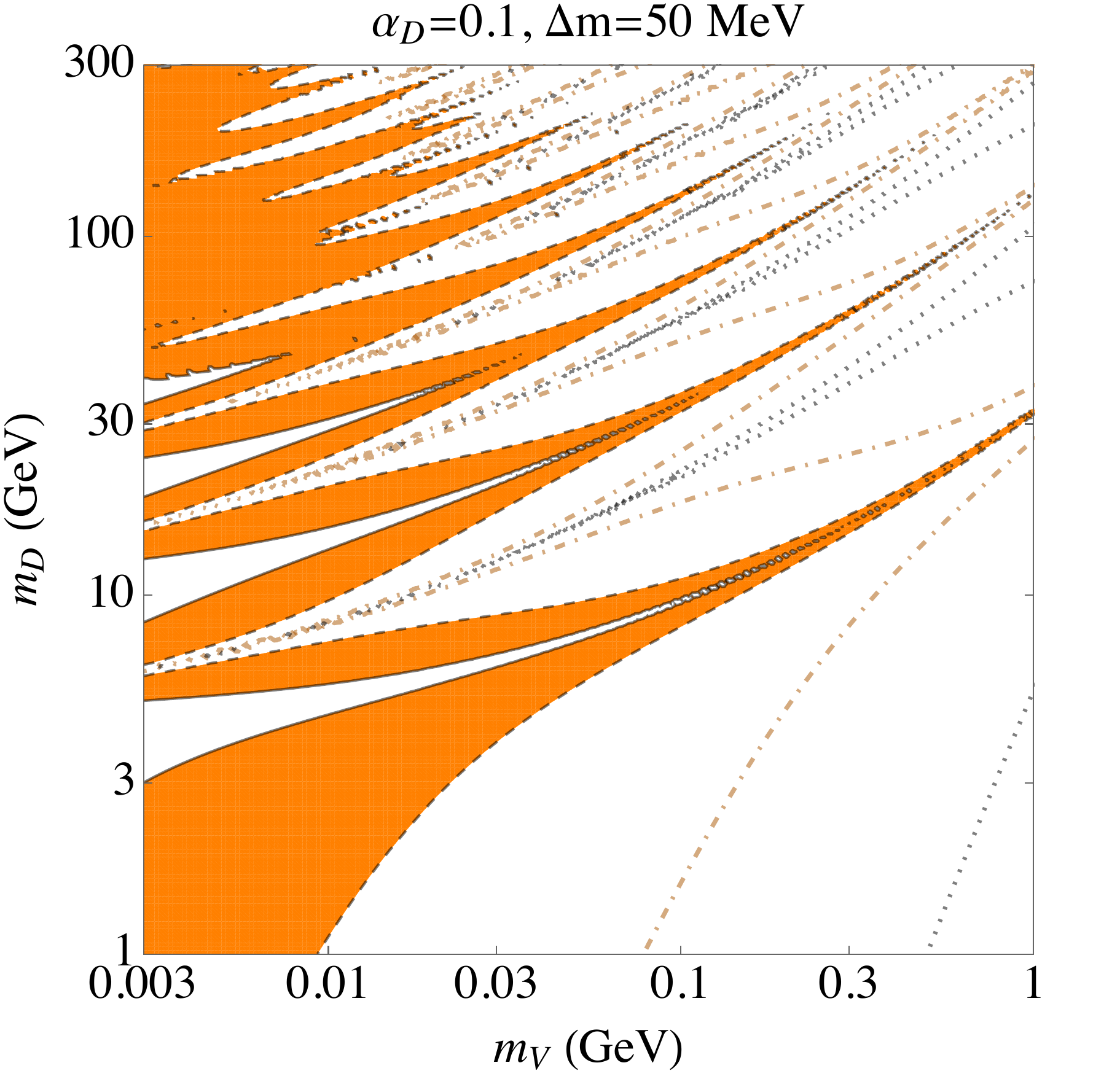}
\includegraphics[width=7.5cm]{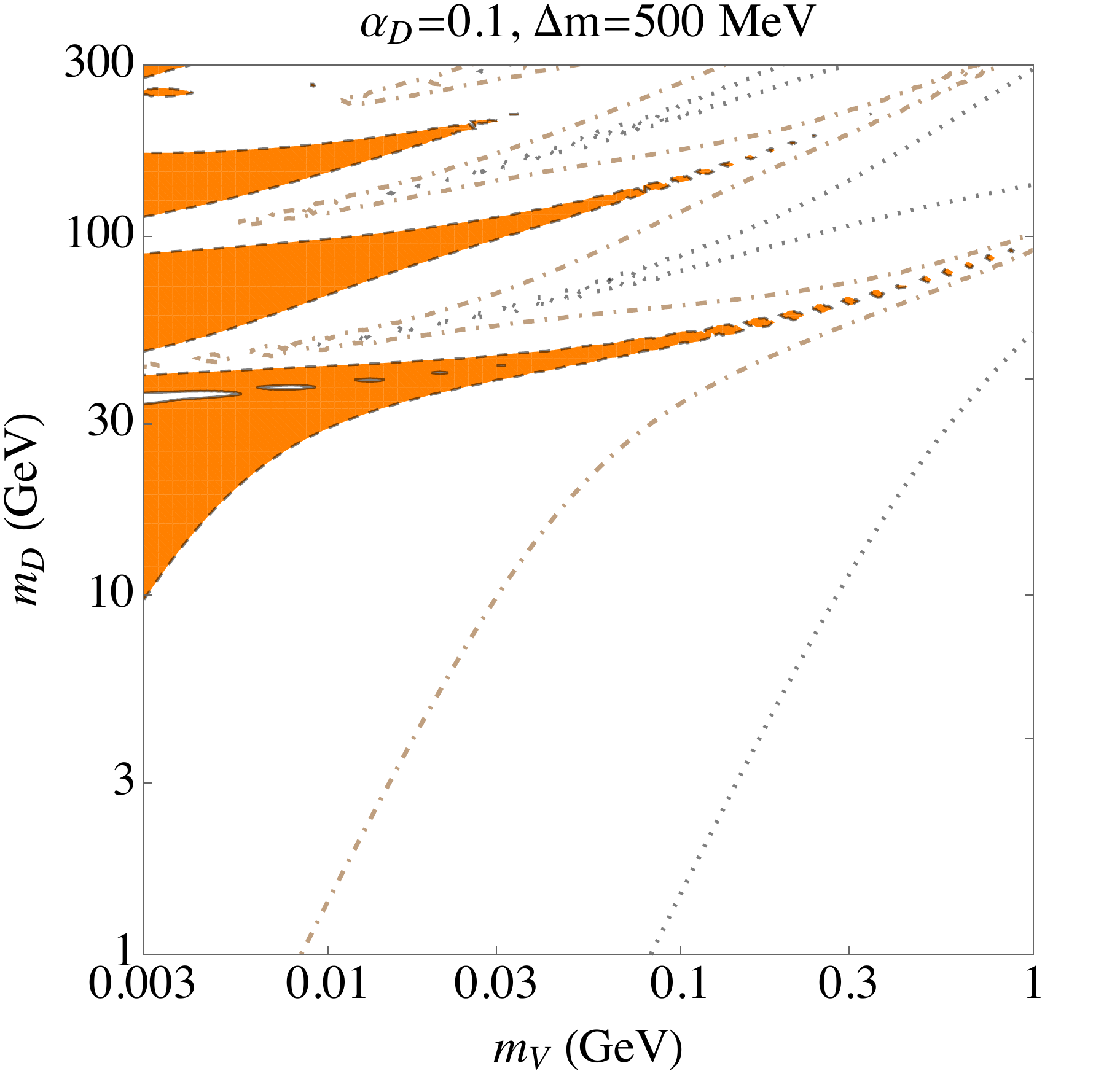}\vspace{3mm}}
\caption{Similar to Fig.~\ref{3}, but with $\alpha_D=0.1$. In this case, larger mass splitting for the dark matter is allowed.}\label{4}
\end{figure}

In principle, one could also vary the values of $\alpha_D$ from what we use here ($10^{-2}$). A smaller $\alpha_D$ would require larger dark matter mass to satisfy the condition $\alpha_D^2 m_D > \Delta m$ thus and further increase the lower bound on SIDM mass.
On the other hand, taking a larger value of $\alpha_D$ than used here could accommodate more SIDM parameter space with a much larger mass splitting $\Delta m$ (see, {\it e.g.}, Fig.~\ref{4} with $\alpha_D=0.1$). 
At the same time the model could receive stronger constraints from dark matter indirect detection, barring a series of astrophysical uncertainties~\cite{Cirelli:2010xx}. 
Very recently, Ref.~\cite{Bringmann:2016din} argued that the cosmic microwave background (CMB) could marginally exclude the parameter space of SIDM. 
However, it is worth noting that \cite{Bringmann:2016din} only works in limit of pure-Dirac SIDM and further assumed thermal production of the dark matter relic density. 
In fact, Ref.~\cite{Slatyer:2009vg} has shown that the CMB constraint could be weakened for pseudo-Dirac dark matter in the presence of an MeV scale mass splitting.
Moreover, if one makes a simple extension to the simple setup considered here to allow the dark photon to decay into other light species in the dark sector (such as the ``dark neutrinos''), 
the CMB bound will be further relaxed. In this sense, the indirect detection constraints are more model dependent.
In contrast, adding such a light dark species will not affect the physics of dark matter self-interaction and direct detection which are involved in the main motivation of this work.

As a reflection on the above results, it may be helpful to consider the annihilation of wino dark matter where a similar quantum mechanical effects play a role~\cite{Hisano:2004ds}. Because of radiative corrections, the charged wino is heavier than the neutral wino dark matter by $166$\,MeV~\cite{Cirelli:2005uq}, which is much larger than the typical dark matter kinetic energy in the galaxy. However, the charged wino still plays an important role in the Sommerfeld enhancement in the dark matter annihilation, because of the potential energy from the exchange of electroweak gauge bosons. Namely, the product of the electroweak coupling strength $\alpha_W^2$ and the TeV wino mass is greater than the charged-neutral wino mass splitting. These enhancements for dark matter annihilation and self-interaction with a large mass splitting share a similar origin.

\section{Conclusion and Outlook}\label{conclusion}

We have explored the case of self-interacting dark matter which aims at solving the small scale structure problems in the context of a dark sector model of massive dark QED. The dark sector couples to the standard model sector via the $U(1)$ kinetic mixing term. We are interested in the question how visible the SIDM from such a dark sector could be probed in our laboratories. Previous calculations on SIDM assumed it is a Dirac fermion in which case the direct detection experiments are at the most frontier of probing the Dirac SIDM except for the low mass (GeV or less) dark matter window. Adding a small Majorana mass to the Dirac fermion dark matter will make it pseudo-Dirac---split it into two nearly-degenerate Majorana mass eigenstates, and if the mass splitting is larger than MeV scale, all direct detection constraints can be evaded. At the same time, as pointed out in this work, the parameter space for SIDM will also be modified. The comparison between the mass splitting $\Delta m$ and the potential energy of order $\sim\alpha_D^2 m_D$ plays an important role in determining the viable parameter space of SIDM. Qualitatively, the parameter space for a pseudo-Dirac self-interacting dark matter without direct detection constraints is $m_D v^2 \ll \Delta m \ll m_D \alpha_D^2$, with $v$ being the local dark matter velocity.

The pseudo-Dirac SIDM case, with the strong direct detection constraints evaded, lights up new hopes of probing the photon-dark-photon kinetic mixing and the dark sector structure through other experiments. The most interesting parameter space is when the dark photon mass lies above $\sim10\,$MeV, where the current high intensity experimental upper bound on the kinetic mixing parameter $\kappa$ is only $\sim10^{-3}$, or even higher. 
Thus the dark photon in the SIDM scenario can still be quite visible with $\kappa$ close to the present upper bound.
There have been several proposals to cover the dark photon in this window~\cite{Echenard:2014lma, Celentano:2014wya, Ilten:2015hya}.
Moreover, if the mass spitting is large enough, according to Fig.~\ref{3} and~\ref{4}, the SIDM mass is pushed up to around the weak scale. The high luminosity and future high energy colliders seem to be very suitable for probing both the SIDM and dark photon in this case. At colliders, the dark matter particles $\chi_1,\, \chi_2$ could be pair-produced via the $U(1)$ kinetic mixing, and dark photon could be radiated from dark matter final states or from the de-excitation of $\chi_2\to\chi_1$. They lead to exotic signatures such as lepton jets~\cite{ArkaniHamed:2008qp, Cheung:2009su, Buschmann:2015awa} and/or displaced vertices~\cite{DeSimone:2010tf, Izaguirre:2015zva}.
For large enough $\alpha_D$ the dark matter bound state channels will also play an important role~\cite{An:2015pva}. We leave a systematic study of these phenomena and their prospects to a future work~\cite{zhen}.

Finally, our work also could be useful when the kinetic mixing vanishes and the dark sector is completely hidden. 
In that case, there is no other option but to continue exploring the nature of dark matter in astrophysical observations.
The new parameter space of pseudo-Dirac SIDM found here is still valid, and is complementary to that for a pure-Dirac dark matter.

\section{Acknowledgement}

I would like to thank Geoffrey Bodwin, Clifford Cheung, Patrick Fox, Andr\'e de Gouv\^ea, Tongyan Lin, Ian Low, Mark Wise and Kathryn Zurek for discussions, and Tongyan Lin for collaboration during the early stages of this work. This work is supported by the DOE grant DE-SC0010143. This work is partly done at the Aspen Center for Physics, which is supported by National Science Foundation grant PHY-1066293.

\end{document}